\documentclass[prb,showpacs,superscriptaddress]{revtex4}

\usepackage{graphicx}
\usepackage{amsmath}
\usepackage{multirow}
\def\fsl#1{\setbox0=\hbox{$#1$}                 % set a box for #1
   \dimen0=\wd0                                 % and get its size
   \setbox1=\hbox{/} \dimen1=\wd1               % get size of /
   \ifdim\dimen0>\dimen1                        % #1 is bigger
      \rlap{\hbox to \dimen0{\hfil/\hfil}}      % so center / in box
      #1                                        % and print #1
   \else                                        % / is bigger
      \rlap{\hbox to \dimen1{\hfil$#1$\hfil}}   % so center #1
      /                                         % and print /
   \fi}                                         %

\newcommand{\Ln}{\mbox{Ln}}

\newcommand{\VEV}[1]{\langle #1 \rangle}

\newcommand{\be}{\begin{equation}}
\newcommand{\ee}{\end{equation}}
\newcommand{\ba}{\begin{eqnarray}}
\newcommand{\ea}{\end{eqnarray}}

\begin{document}

\title{Toward theory of quantum Hall effect in graphene}

\date{\today}

%\preprint{UWO-TH-07/}

\author{E.V. Gorbar}
\affiliation{Bogolyubov Institute for Theoretical
Physics, 03680, Kiev, Ukraine}

\author{V.P. Gusynin}
\affiliation{Bogolyubov Institute for Theoretical
Physics, 03680, Kiev, Ukraine}

\author{V.A. Miransky}
\altaffiliation[On leave from ]{Bogolyubov Institute for
Theoretical Physics, 03680, Kiev, Ukraine}

\affiliation{Department of Applied Mathematics, University of Western Ontario,
London, Ontario N6A 5B7, Canada}

%\date{\today}

\begin{abstract}

We analyze a gap equation for the propagator of Dirac quasiparticles
and conclude that in
graphene in a magnetic field, the order parameters connected with the quantum
Hall ferromagnetism dynamics and those connected with the magnetic catalysis
dynamics necessarily coexist (the latter have the form of Dirac masses and
correspond to excitonic condensates). This feature of graphene could lead to
important consequences, in particular, for the existence of gapless edge
states. Solutions of the gap equation corresponding to recently experimentally
discovered novel plateaus in graphene in strong magnetic fields are described.

\end{abstract}


\maketitle

The properties of graphene, a single atomic layer of graphite
\cite{Geim2004Science}, have attracted great interest, especially after the
experimental discovery \cite{Geim2005Nature,Kim2005Nature} and (made
independently) theoretical prediction \cite{Ando2002,Gusynin2005PRL,Peres2005}
of an anomalous quantization in the quantum Hall (QH) effect. In this case, the
filling factors are $\nu = \pm 4(|n| + 1/2)$, where $n$ is the Landau level
index. For each QH state, a four-fold (spin and sublattice-valley) degeneracy
takes place. These properties of the QH effect are intimately connected with
relativistic like features in the graphene dynamics
\cite{Semenoff1984PRL,Haldane1988PRL,Luk'yanchuk2004}.

In recent experiments \cite{Zhang2006,Jiang2007}, it has been observed that in
a strong enough magnetic field, $B \gtrsim 20\, \mbox{T}$, the new QH plateaus,
$\nu = 0, \pm 1$ and $\pm 4$, occur, that was attributed to the magnetic field
induced splitting of the $n =0$ and $n = \pm 1$ Landau levels (LLs). It is
noticeable that while the degeneracy of the lowest LL (LLL), $n = 0$, is
completely lifted, only the spin degeneracy of the $n = \pm 1$ LL is removed.

On theoretical side, there are now two leading scenarios for the
description of these plateaus. One of them is the QH ferromagnetism (QHF)
\cite{Nomura2006PRL,Goerbig2006,Alicea2006PRB,Sheng2007} (the dynamics of a
Zeeman spin splitting enhancement considered in Ref. \cite{Abanin2006PRL} is
intimately connected with the QHF). The second one is the magnetic catalysis
(MC) scenario in which excitonic condensates (Dirac masses) are spontaneously
produced \cite{Gusynin2006catalysis,Herbut2006,Fuchs2006,Ezawa2006}. For a
brief review of these two scenarios, see Ref. \cite{Yang2007}.

While the QHF scenario is based on the dynamical framework developed for
bilayer QH systems \cite{Arovas1999}, the MC scenario is based on the
phenomenon of an enhancement of the density of states in a strong magnetic
field, which catalyzes electron-hole pairing (leading to excitonic condensates)
in relativistic like systems. The essence of this effect is the dimensional
reduction $D \to D - 2$ in the electron-hole pairing dynamics and the presence
of the LLL with energy $E = 0$ (containing both electron and hole states) in
relativistic systems in a magnetic field. This universal phenomenon was
revealed in Ref. \cite{Gusynin1995PRD} and was first considered in graphite in
Refs. \cite{Khveshchenko2001PRL,Gorbar2002PRB}.

On technical side, the difference between these two scenarios is in utilizing
different order parameters in breaking the spin-sublattice-valley $U(4)$
symmetry of the noninteracting Hamiltonian of graphene. While the QHF order
parameters are described by densities of the conserved charges connected with
diagonal generators of the non-abelian subgroup $SU(4) \subset U(4)$, the order
parameters in the MC scenario are Dirac mass like terms. Note that while the
latter are bifermion operators which are invariant under 2 + 1 dimensional
Lorentz transformations (with the Fermi velocity $v_F \simeq 10^6 \mbox{m/s}$
playing the role of light velocity), the QHF charge densities are time like
components of the corresponding conserved currents which transform as vectors
under the Lorentz transformations.

One may think that the QHF and MC order parameters should compete
with each other. However, as will be shown in this paper, the
situation is quite different: These two sets of the order parameters
{\it necessarily} coexist, which implies that they have the same
dynamical origin. The physics underlying their coexistence is
specific for relativistic like dynamics that makes the QH dynamics
of the $U(4)$ breakdown in graphene to be quite different from that
in bilayer QH systems \cite{Arovas1999} whose dynamics have no
relativistic like features.

Our approach is based on studying the gap equation for the propagator
of Dirac quasiparticles.
For the description of the dynamics in graphene, we will use the same model as
in Refs. \cite{Khveshchenko2001PRL,Gorbar2002PRB}, in which while
quasiparticles are confined to a 2-dimensional plane, the electromagnetic
(Coulomb) interaction between them is three-dimensional in nature. The dynamics
will be treated in the Hartree-Fock (mean field) approximation, which is
conventional and appropriate in this case \cite{Khveshchenko2001PRL,
Gorbar2002PRB,Nomura2006PRL,Goerbig2006,Gusynin2006catalysis}. Then, at zero
temperature and in the clean limit (no impurities), the gap equation takes the
form:
\begin{eqnarray}
&&\hspace{-5mm}G^{-1}(x,y) = S^{-1}(x,y) + i\hbar\gamma^0G(x,y)\gamma^0
\delta(x_0 -
y_0)U_{C}(\vec{x}-\vec{y}) \nonumber\\
&&- i\hbar\gamma^0 \mbox{tr}[\gamma^0G(x,x)]\delta^{3}(x-
y)U^{F}_{C}(0). \label{SD}
\end{eqnarray}
Here $x \equiv (x_0,\vec{x})$, with $x_0 \equiv t$ being time coordinate,
$U_{C}(\vec{x})$ is the Coulomb potential in a magnetic field, given in Eq.
(46) in Ref. \cite{Gorbar2002PRB}, $U^{F}_{C}(0)$ is its Fourier transform at
$\mathbf{k}=0$, $G^{-1}(x,y)$ is the full inverse
quasiparticle
propagator, and $S^{-1}(x,y)$ is the bare inverse quasiparticle propagator,
\begin{equation}
iS^{-1}(x,y)={[(i\hbar\partial_t+\mu_0 - \mu_BB\sigma^3)\gamma^0 -
v_{F}\vec{\pi}\vec{\gamma}]\delta^{3}(x-y)}, \label{inversebare}
\end{equation}
where  $\mu_0$ is the electron chemical potential,
$\vec{\pi}=-i\hbar\vec{\partial}+e\vec{A}/c$ is the canonical momentum, and
$\mu_BB\gamma^0\sigma^3$ is the Zeeman term [the vector potential $\vec{A}$
corresponds to the magnetic field $\vec{B}$, $B \equiv |\vec{B}|$,
$\mu_B$ is the Bohr magneton, and the
Pauli matrix $\sigma^3$ acts on spin indices] \cite{Zeeman}.

For Dirac matrices $\gamma^0, \vec{\gamma}$, we use the same representation as
in Refs. \cite{Gorbar2002PRB,Gusynin2006catalysis} ($xy$-plane is chosen for
graphene). Note that while the second term on the right hand side of
Eq.(\ref{SD}) describes exchange interactions, the third one is the
Hartree term describing annihilation interactions.

The analysis of gap equation (\ref{SD}) beyond the LLL approximation is a very
formidable problem. Because of that, we will utilize the following
approximation: the Coulomb potential $U_{C}(\vec{x})$ in the gap equation will
be replaced by the contact interaction $G_{int}\delta^{2}(\vec{x})$:
\begin{eqnarray}
G^{-1}(x,y) = S^{-1}(x,y)
+ i\hbar G_{int}\gamma^0G(x,x)\gamma^0 \delta^{3}(x - y)
- i\hbar G_{int}\gamma^0\, \mbox{tr}[\gamma^0G(x,x)]\delta^{3}(x- y),
\label{gap}
\end{eqnarray}
where $G_{int}$ is a dimensional coupling constant. Such an
approximation is common in Quantum Chromodynamics (QCD), where long
range gluon interactions are replaced by contact
(Nambu-Jona-Lasinio) ones. This leads to a good description of
nonperturbative dynamics in low energy region in QCD (for a review,
see for example Ref. \cite{book}). Because of the universality of
the MC phenomenon and because the symmetric and kinematic structures
of the gap equations (\ref{SD}) and (\ref{gap}) are the same, we
expect that approximate gap equation (\ref{gap}) should be at least
qualitatively reliable for the description of the LLL and first few
LLs, say, $n=\pm 1$ LL. This in turn implies
that in the analysis of this gap equation, one should use an
ultraviolet cutoff $\Lambda$ of the order of the Landau scale
$L(B)\equiv\sqrt{\hbar|eB_{\perp}|v_{F}^2/c}\simeq 300
\sqrt{B_{\perp}({\rm T})}\, [{\rm K}]$ (in Kelvin), where
$B_{\perp}$ is the component of $\vec{B}$ orthogonal to the graphene
plane measured in Tesla. The dimensional coupling constant $G_{int}$
should be taken then as $G_{int} \sim 1/\sqrt{eB_\perp}$ (see
below) \cite{Katsnelson2006}.

Because of the Zeeman term, the $U(4)$ symmetry is broken down to the
``flavor'' symmetry $U(2)_{+} \times U(2)_{-}$, where the subscript $\pm$
corresponds to up and down spin states respectively. The generators of the
$U(2)_{s}$, with $s=\pm$,
are $I \otimes P_s$, $-i\gamma^3 \otimes P_s$, $\gamma^5 \otimes
P_s$, and $\gamma^3\gamma^5 \otimes P_s$ (here $I$ is the $4 \times 4$ unit
matrix, $\gamma^5 = i\gamma^0\gamma^1\gamma^2\gamma^3$, and $P_{\pm}=(1 \pm
\sigma^3)/2$ are projectors on spin up and spin down states)
\cite{Gusynin2006catalysis}.

Our goal is searching for solutions of Eq. (\ref{gap}) both with spontaneously
broken and unbroken $SU(2)_{s}$, where $SU(2)_{s}$ is the largest non-abelian
subgroup of the $U(2)_{s}$. The Dirac mass term
$\tilde{\Delta}_s\bar{\psi}P_{s}\psi \equiv
\tilde{\Delta}_s\psi^{\dagger}\gamma^{0}P_{s}\psi$,
where $\tilde{\Delta}_s$ is a Dirac
gap (mass), is assigned to the triplet representation of the
$SU(2)_{s}$, and the generation of such a mass would lead to spontaneous flavor
$SU(2)_{s}$ symmetry breaking down to the $\tilde{U}(1)_{s}$ with the generator
$\gamma^3\gamma^5 \otimes P_s$
\cite{Gusynin2006catalysis,Khveshchenko2001PRL,Gorbar2002PRB}. There is also a
Dirac mass term of the form
$\Delta_s\bar{\psi}\gamma^3\gamma^{5}P_s\psi$
that is a singlet with respect to $SU(2)_{s}$, and therefore its generation
would not break this symmetry. On the other hand, while the triplet mass term
is even under time reversal $\cal{T}$, the singlet mass term is $\cal{T}$-odd
(for a recent review of the transformation properties of different mass terms
in graphene, see Ref. \cite{Gusynin2007review}). It is noticeable that
consequences of the presence of the mass $\Delta$ in graphite were discussed
long ago in Ref. \cite{Haldane1988PRL}.

The analysis of gap equation (\ref{gap}) that we use is closely connected with
that in Ref. \cite{Gusynin1995PRD} and based on the decomposition of the
quasiparticle propagator over the LL poles with the residues expressed
through the
generalized Laguerre polynomials. A detailed description of the analysis will
be presented elsewhere. Here we will describe its main results. It was found
that, for a fixed spin, the full inverse quasiparticle propagator takes the
following general form (compare with Eq. (\ref{inversebare})):
\begin{eqnarray}
iG^{-1}_{s}(x,y)= [(i\hbar\partial_t+\mu_{s} +\tilde{\mu}_s\gamma^3\gamma^5)\gamma^0
- v_{F}\vec{\pi}\vec{\gamma}
- \tilde{\Delta}_s + \Delta_s\gamma^3\gamma^5]\delta^{3}(x-y),
\label{full-inverse}
\end{eqnarray}
where the parameters $\mu_s$, $\tilde{\mu}_s$, $\Delta_s$, and
$\tilde{\Delta}_s$ are determined from gap equation (\ref{gap}). Note that the
chemical potential $\mu_\pm$ includes the Zeeman energy $\mp Z$,
with $Z = \mu_B B = 0.67 B({\rm T})\, [{\rm K}]$, and
the chemical potential $\tilde{\mu}_s$ is related to the density of the
conserved pseudospin charge
$\psi^{\dagger}\gamma^3\gamma^{5}P_s\psi$, which is assigned to the triplet
representation of the $SU(2)_{s}$. Therefore, while the masses $\Delta_s$ and
$\tilde{\Delta}_s$ are related to the MC order parameters
$\VEV{\bar{\psi}\gamma^3\gamma^{5}P_s\psi}$ and
$\VEV{\bar{\psi}P_s\psi}$,
the chemical potentials $\mu_3 \equiv (\mu_+ -
\mu_-)/2$ and $\tilde{\mu}_s$ are related to the conventional QHF ones: the
spin density $\VEV{\psi^{\dagger}\sigma^3 \psi}$ and the pseudospin density
$\VEV{\psi^{\dagger}\gamma^3\gamma^5P_s\psi}$, respectively. Note
that while the triplet Dirac mass term describes the charge density
imbalance between the two graphene sublattices
\cite{Khveshchenko2001PRL,Gusynin2006catalysis}, the pseudospin density
describes the charge density imbalance between the two
valley points in the Brillouin zone.

The dispersion relations
for higher LLs ($|n| \geq 1$) following from Eq. (\ref{full-inverse})
are
\begin{eqnarray}
\hspace{-2mm}E^{(\sigma)}_{ns} =-\mu_s +\sigma\tilde{\mu}_s
+\mbox{sign}(n)\sqrt{2\hbar |neB_{\perp}|v_{F}^2/c + (\tilde{\Delta}_s
+\sigma\Delta_s)^2}\,\,,
\label{higherLLs}
\end{eqnarray}
where $\sigma=\pm 1$ are connected with eigenvalues of the pseudospin matrix
$\gamma^3\gamma^5$.
The case of the LLL is special, and its dispersion
relation is
\begin{equation}
E^{(\sigma)}_{s}= -\mu_s + \sigma[\tilde{\mu}_s\,\mbox{sign}(eB_{\perp})
+\,\tilde{\Delta}_s] + \Delta_s\,\mbox{sign}(eB_{\perp}).
\label{LLLenergylevels}
\end{equation}
One can see from Eqs. (\ref{higherLLs}),
(\ref{LLLenergylevels})
that at a fixed spin,
the terms with $\sigma $ are responsible
for splitting of LLs.

In fact, for each value of spin, our analysis revealed the following three
types of solutions: a) a singlet solution with a nonzero singlet mass $\Delta$
and with no triplet parameters $\tilde{\Delta}$ and $\tilde{\mu}$, b) a triplet
solution with nonzero $\tilde{\Delta}$ and $\tilde{\mu}$, and with the singlet
mass $\Delta$ being zero, and c) a mixed solution with $\Delta$,
$\tilde{\Delta}$, and $\tilde{\mu}$ being nonzero. The latter is realized only
in higher LLs. In order to find the most stable solution among them, we compare
the free energy density $\Omega$ of the corresponding ground states. In the
mean field approximation that we use, $\Omega$ takes the following form on
solutions of the gap equation \cite{potential}
\be
\Omega VT=i{\rm Tr}\left[\Ln
G^{-1} +\frac{1}{2}\left(S^{-1}G-1\right)\right], \label{potential}
\ee
where $VT$ is the space-time volume,
the trace, the logarithm, and the product $S^{-1}G$ are taken in the functional
sense, and $G = \rm{diag}(G_{+}, G_{-})$.

The process of filling the LLs is described by varying the electron
chemical potential $\mu_0$. We will consider positive $\mu_0$
(dynamics with negative $\mu_0$ is related by electron-hole symmetry
and will not be discussed separately).
In this paper we will mostly consider the LLL dynamics (results for
the $n = 1$ LL will be briefly described at the end of the paper).

For the case when only the LLL is doped, which corresponds to the condition
$|\mu_s \pm \tilde{\mu}_s| \ll L(B)$,
we arrive at the following results:

i) A solution with singlet Dirac masses both for spin up and spin down is the
most favorable for $0 \le \mu_0 < 2A+Z$, where $A \equiv
G_{int}|eB_\perp|/8\pi\hbar c$\, \cite{footnote3}.
It is:
\begin{equation}
\tilde{\Delta}_\pm= \tilde{\mu}_\pm=0,\,\, \mu_\pm  =\bar{\mu}_{\pm} \mp A,\,\,
\Delta_\pm = \pm M\,\mbox{sign}(eB_\perp) \label{singlet}
\end{equation}
with $\bar{\mu}_{\pm} \equiv \mu_0 \mp Z$ and $M \equiv A/(1- \lambda)$,
$A=\lambda\sqrt{\pi}L^{2}(B)/2\Lambda$ where
the dimensionless coupling constant $\lambda$ is $\lambda \equiv
G_{int}\Lambda/(4\pi^{3/2}\hbar^{2}v_{F}^2)$ \cite{footnote4}. From dispersion
relation (\ref{LLLenergylevels}), we find that $E_+ > 0$ and $E_- < 0$, i.e.,
the LLL is half filled (the energy spectrum in this solution is $\sigma$
independent). Therefore the spin gap $\Delta E_{0} = E_+ - E_-$ corresponds to
the $\nu = 0$ plateau. The value of the gap is $\Delta E_{0} = 2M + 2(Z + A)$.
It is instructive to compare $\Delta E_{0}$ with the spin gap in Ref.
\cite{Abanin2006PRL}. The latter contains an enhanced Zeeman spin splitting,
which corresponds to the second term $2(Z + A)$ in $\Delta E_{0}$. However,
besides this term, there is also the large contribution $2M$ in $\Delta E_{0}$
in the present solution, which is connected with a dynamical singlet Dirac
mass for quasiparticles. The presence of this mass could have important
consequences for gapless edge states whose relevance for the physics of the
$\nu = 0$ plateau was pointed out in Ref. \cite{Abanin2007PRL}.
Generalizing the analysis in \cite{Abanin2007PRL}, we have found
that such states exist only when the full Zeeman splitting $Z + A$
is larger than the Dirac gap $M=A/(1-\lambda)$. This leads to the constraint
$Z > \lambda A/(1 - \lambda)$. Let us consider the case with $B = B_\perp$.
Then, since $Z \sim B_{\perp}$ and $A \sim \sqrt{B_{\perp}}$ (see below), this
constraint leads to a lower limit $B_{\perp}^{(cr)}$ for the values of
$B_{\perp}$ at which gapless edge states exist. On the other hand, since $Z$
depends on total $B$ while $A$ depends only on $B_\perp$, adding a longitudinal
$B_{||}$ will decrease the lower limit for $B_\perp$. It would be interesting
to check experimentally this point.
Also, these features could be
relevant for the interpretation of
the recent experiments \cite{Ong2007}, in which
no gapless edge states were detected for $B=B_{\perp} \le 14\,{\rm T}$.
We shall return to this issue below.

ii) A hybrid solution, with a triplet Dirac mass for spin up and a singlet
Dirac mass for spin down, is the most favorable for $2A+Z \le \mu_0 < 6A+Z$. It
is:
$$
\tilde{\Delta}_{+} = M,\,\,\, \tilde{\mu}_{+} = A\,\mbox{sign}(eB_\perp)
,\,\,\, \mu_{+} = \bar{\mu}_+ - 4A,\,\,\,\, \Delta_+ =0,
$$
\begin{eqnarray}
\tilde{\Delta}_-=\tilde{\mu}_-=0,\,\, \mu_- = \bar{\mu}_- -3A,\,\, \Delta_- =
-M\,\mbox{sign}(eB_\perp). \label{hybrid}
\end{eqnarray}
As follows from Eq. (\ref{LLLenergylevels}), while $E_{+}^{(+1)} > 0$, the
energies $E_{+}^{(-1)}$ and $E_{-}^{(+1)}= E_{-}^{(-1)}$ are negative.
Consequently, the LLL is now three-quarter filled and, therefore, the gap
$\Delta E_{1} = E_{+}^{(+1)} - E_{+}^{(-1)} = 2(M + A)$ corresponds to the $\nu
= 1$ plateau. The latter, unlike the $\nu = 0$ plateau, is directly related to
spontaneous $SU(2)_{+}$ flavor symmetry breaking.

iii) A solution with equal singlet Dirac masses for spin up and spin down
states is the most favorable for $\mu_0 > 6A+Z$. It is
\begin{eqnarray}
\tilde{\Delta}_\pm =\tilde{\mu}_\pm=0,\, \mu_\pm =\bar{\mu}_{\pm} - 7A,
\Delta_\pm =-M \mbox{sign}(eB_\perp) \label{singlet1}
\end{eqnarray}
(compare with Eq. (\ref{singlet})). It is easy to check from
(\ref{LLLenergylevels}) that both $E_+$ and $E_-$ are negative in this case,
i.e., the LLL is completely filled. Therefore,
this solution corresponds to the $\nu = 2$
plateau related to the energy gap $\Delta E_2 \simeq \sqrt{2}L(B)$
between the LLL and the $n=1$ LL.

 %Finally, we checked that although the parameters of the solutions
%i)-iii) abruptly jump at the points $\mu_0 = 2A+Z$ and
%$\mu_0 = 6A+Z$, the free
%energy densities of the corresponding solutions match exactly. Therefore
%typical first order phase transitions take place at these values of the
%electron chemical potential $\mu_0$.

This analysis leads us to the picture for the LLL plateaus which qualitatively
agrees with that in experiments \cite{Zhang2006,Jiang2007}. In particular,
taking the dimensionless coupling $\lambda$ to be a free parameter and choosing
cutoff $\Lambda$ to be of the order of the Landau scale $L(B)$, we arrive at the scaling relations,
$A  \sim \sqrt{|eB_{\perp}|}$, $M \sim \sqrt{|eB_{\perp}|}$, and, therefore, $\Delta
E_1 = 2(A + M) \sim \sqrt{|eB_{\perp}|}$ for the gap related to the $\nu = 1$
plateau. One can check that the experimental value $\Delta E_1
\sim 100\,{\rm K}$
for $B_{\perp}= 30\,{\rm T}$ \cite{Jiang2007} corresponds to
$\lambda \sim 0.02$.
However, because interactions with impurities are ignored in the clean
limit used in the present model,
it would be more reasonable to consider
$\lambda$, say, in interval $0.02 - 0.2$. Then, for these values of $\lambda$,
we find from the constraint $Z > \lambda A/(1 - \lambda)$ in
the solution i) above
that the gapless edge states exist for $|B_\perp| > B_{\perp}^{(cr)}$, where
$0.01\, {\rm T} \alt B_{\perp}^{(cr)} \alt 200\,{\rm T}$. One can see that
$B_{\perp}^{(cr)}$ is sensitive to the choice of $\lambda$.
Therefore in order to fix the
critical value $B_{\perp}^{(cr)}$ more accurately, one should utilize a
more realistic and constrained model.

As to the $n=1$ LL, we found that there are the gaps
$\Delta E_{3}=\Delta E_{5} \simeq 2A$ and $\Delta E_{4}\simeq 2(Z+A)$
corresponding to the plateaus $\nu=3, 5$ and $\nu=4$, respectively
(the contributions of Dirac masses are suppressed
at least by factor $M^{2}/L^{2}(B)$ there). Note that $\Delta E_{3,5}$
and $\Delta E_{4}$ are essentially smaller than the LLL gaps
$\Delta E_{1}$ and
$\Delta E_{0}$, respectively ($\Delta E_{3,5}\alt \Delta E_{1}/2$).
On the other hand,
the experimental data yield
$\Delta E_{4}\simeq 2Z$, and no gaps $\Delta E_{3}, \Delta E_{5}$
have been observed \cite{Zhang2006,Jiang2007}.
We believe that a probable explanation of this
point is that, unlike $Z$,
the value of the {\it dynamically} generated parameter
$A$ corresponding to the $|n| \geq 1$ LLs will be essentially
reduced if a considerable broadening of higher LLs in a magnetic field
is taken into account \cite{Gusynin2006catalysis}.
If so, the gap $\Delta E_{4}$ will be reduced to $2Z$
and the gaps
$\Delta E_{3}, \Delta E_{5}$ will become unobservable.

Recently, in Ref. \cite{Giesbers2007},
a large width $\Gamma_1$ of $400\,{\rm K}$ was determined for the $n=1$ LL.
The plateaus $\nu =3, 5$
could become observable if
the gaps $\Delta E_3 =\Delta E_5 \simeq 2A$ calculated in
the clean limit are at least of order $\Gamma_1$ or larger
\cite{Gusynin2006catalysis} .
The LLL gap $\Delta E_1 \simeq 100\,{\rm K}$ at $|B_\perp| = 30\,{\rm T}$
corresponds to $\Delta E_{3,5} \alt 50\,{\rm K}$.
Then, taking a conservative estimate $\Gamma_1 = 100\,{\rm K}$ and
using $A \sim \sqrt{|eB_\perp|}$, we conclude that to
observe the $\nu =3, 5$ plateaus, the magnetic fields should be
at least as large as $B \sim 100\,{\rm T}$.

In conclusion, we have shown that the QHF and MC order parameters
in graphene are two sides of the same coin and they necessarily coexist.
This feature
could have important dynamical consequences for low energy excitations,
in particular, for gapless edge
states. It would be desirable to extend the present analysis to a more
realistic model setup, including the genuine Coulomb interactions, LLs impurity
scattering rates, and temperature.

Useful discussions with S.G. Sharapov and I.A. Shovkovy are acknowledged. The
work of E.V.G and V.P.G. was supported by the SCOPES-project IB 7320-110848 of
the Swiss NSF, the grant 10/07-H "Nanostructure systems, nanomaterials,
nanotechnologies", and by the Program of Fundamental Research of the Physics
and Astronomy Division of the National Academy of Ukraine. V.A.M. acknowledges
the support of the Natural Sciences and Engineering Research Council of Canada.


\begin{thebibliography}{99}
\bibitem{Zeeman} The Zeeman coupling $\mu_{B}\vec{B}\vec{\sigma}$
can be always diagonalized in the spin space as $\mu_{B}B\sigma^3$.
\bibitem{footnote3}
In dynamics in a magnetic field at zero temperature, there
is no one-to-one correspondence between electron density
and chemical potential. As a result, different values of the
latter may correspond to the same physics, as it takes
place for this solution.
\end{thebibliography}
\end{document}